\documentclass[journal=jacsat,manuscript=article]{achemso}
\usepackage[version=3]{mhchem} 
\usepackage{hyperref}
\usepackage{pdfpages} 

\author{Danielle N. Alverson}
\affiliation{Department of Materials Science and Engineering, University of Florida, Gainesville, FL 32611, USA}

\author{Daniel Olds}
\affiliation{National Synchrotron Light Source II, Brookhaven National Laboratory, Upton, NY 11973, USA}

\author{Megan M. Butala}
\affiliation{Department of Materials Science and Engineering, University of Florida, Gainesville, FL 32611, USA}
\email{mbutala@ufl.edu}
\title{Distinguishing Isotropic and Anisotropic Signals for X-ray Total Scattering using Machine Learning}

\begin{document}

\begin{abstract}
Understanding structure$-$property relationships is essential for advancing technologies based on thin film materials. X-ray pair distribution function (PDF) analysis can access relevant atomic structure details spanning local-, mid-, and long-range order. While X-ray PDF has been adapted for thin films on amorphous substrates, measurements of films on single crystal substrates are necessary for accurately determining structure origins for some thin film materials, especially those for which the substrate changes the accessible structure and properties. However, when measuring thin films on single crystal substrates, high intensity anisotropic Bragg spots saturate 2D detector images, overshadowing the thin films' isotropic scattering signal. This renders previous data processing methods, developed for films on amorphous substrates, unsuitable for films on single crystal substrates. To address this measurement need, we developed IsoDAT2D, an innovative data processing approach using unsupervised machine learning algorithms. The program combines non-negative matrix factorization and hierarchical agglomerative clustering to effectively separate thin film and single crystal substrate X-ray scattering signals. We used SimDAT2D, a program we developed to generate synthetic thin film data, to validate IsoDAT2D, and also successfully use the program to isolate X-ray total scattering signal from a thin film on a single crystal Si substrate. The resulting PDF data are compared to similar data processed using previous methods, demonstrating superior performance relative to substrate subtraction with a single crystal substrate and similar performance to substrate subtraction from a film on an amorphous substrate. With IsoDAT2D, there are new opportunities to expand the use of PDF to a wider variety of thin films, including those on single crystal substrates, with which new structure$-$property relationships can be elucidated to enable fundamental understanding and technological advances.

\end{abstract}

\maketitle

\section{Introduction}

Amorphous, nanocrystalline, and otherwise disordered thin film materials have distinct properties from their bulk states that can be beneficial in applications such as spintronics,\cite{bauer_dysprosium_2020} phase change memory,\cite{scott_thermal_2020, raoux_phase_2014} photonics,\cite{Ross_2016} and broadly in semiconductor devices.\cite{Nishida_2018} The ever-increasing demand for high-performance computers and advanced electronic technologies necessitates the continual improvement of the materials that drive them.\cite{Chris_Semiconductor} Atomic structure models of these materials are essential to enhance their properties, optimize their performance, and advance our fundamental understanding of the origins of compelling functional properties. However, current atomic structure characterization is limited for these types of thin films, specifically for local- and mid-range atomic structure features.

Part of the challenge arises from the relative volume of a thin film and its substrate. Thin films are typically deposited or grown on a substrate via chemical or physical vapor deposition processes,\cite{dapkus1982metalorganic,pvd_epitaxial_Arndt} resulting in films with thicknesses between 1\,nm and several microns.\cite{Ude_2018,kotsonis2020property, Frances_2021} The thin films intended for practical applications are typically deposited on single crystal substrates, such as Si, perovskites, and sapphire. This material selection for the substrate can affect the structure and properties of the films. These structure$-$property relationships can be tuned directly for implementation into semiconductor manufacturing processes.\cite{Im_James_1997, Seo_ovonics_2022, P_A_Ivanov_1992,McLellan2023, Rosenberg2021, bauer_dysprosium_2020} Substrates are typically $\approx$500\,$\mu$m thick single crystal wafers.\cite{Will_SiC_Si_1983, Zhao_2004, Matienzo_1993} As substrates are the majority of the volume of thin film samples, isolating the properties and structure of the relatively small volume of the film is complex.

X-ray diffraction (XRD) is a primary technique for determining the atomic structure of crystalline films, which is often used alone or in combination with transmission electron microscopy, especially for epitaxial films.\cite{Hsu_Raj_1992,osherov_role_2007,MoreiradosSantos2003, Kim2013} However, for amorphous or disordered materials, traditional crystallographic methods, such as XRD, can miss local structure motifs that are uncorrelated at long ranges and not captured in an average structure model.\cite{Egami_L._2003, laurita2017chemical} By using only Bragg scattering, XRD analysis excludes information from diffuse scattering and is unsuitable for probing structural features of amorphous materials.\cite{Ohara_2021,playford2014new} 

The pair distribution function (PDF), a sine Fourier transform of total scattering data, uniquely provides local-, mid-, and long-range atomic structural information, even for disordered materials.\cite{Butala_2018,Page_2016,chepkemboi2022strategies} During PDF measurements of powdered samples, scattering data are collected over a wide range of momentum transfer, $Q$, in transmission geometry. This requires short sample-to-detector distances and high-energy radiation, typically using synchrotron X-ray or spallation neutron sources.\cite{Egami_L._2003,proffen_structural_2003, Sivia2011}

Billinge and collaborators demonstrated that powder methods of X-ray total scattering, \textit{i.e.}, acquiring data in transmission geometry, could also be applied to thin film materials on amorphous substrates.\cite{jensen_demonstration_2015} This involves collecting total scattering data from a thin film sample on a substrate and a substrate without a film and subtracting the latter from the former to isolate the thin film signal.\cite{jensen_demonstration_2015,Dippel:ro5016, johnson_david_2017} The total scattering data for the film is then Fourier transformed to yield the PDF. 

\begin{figure}
\graphicspath{ {./Figures/} }
\includegraphics[width = 3.5in]{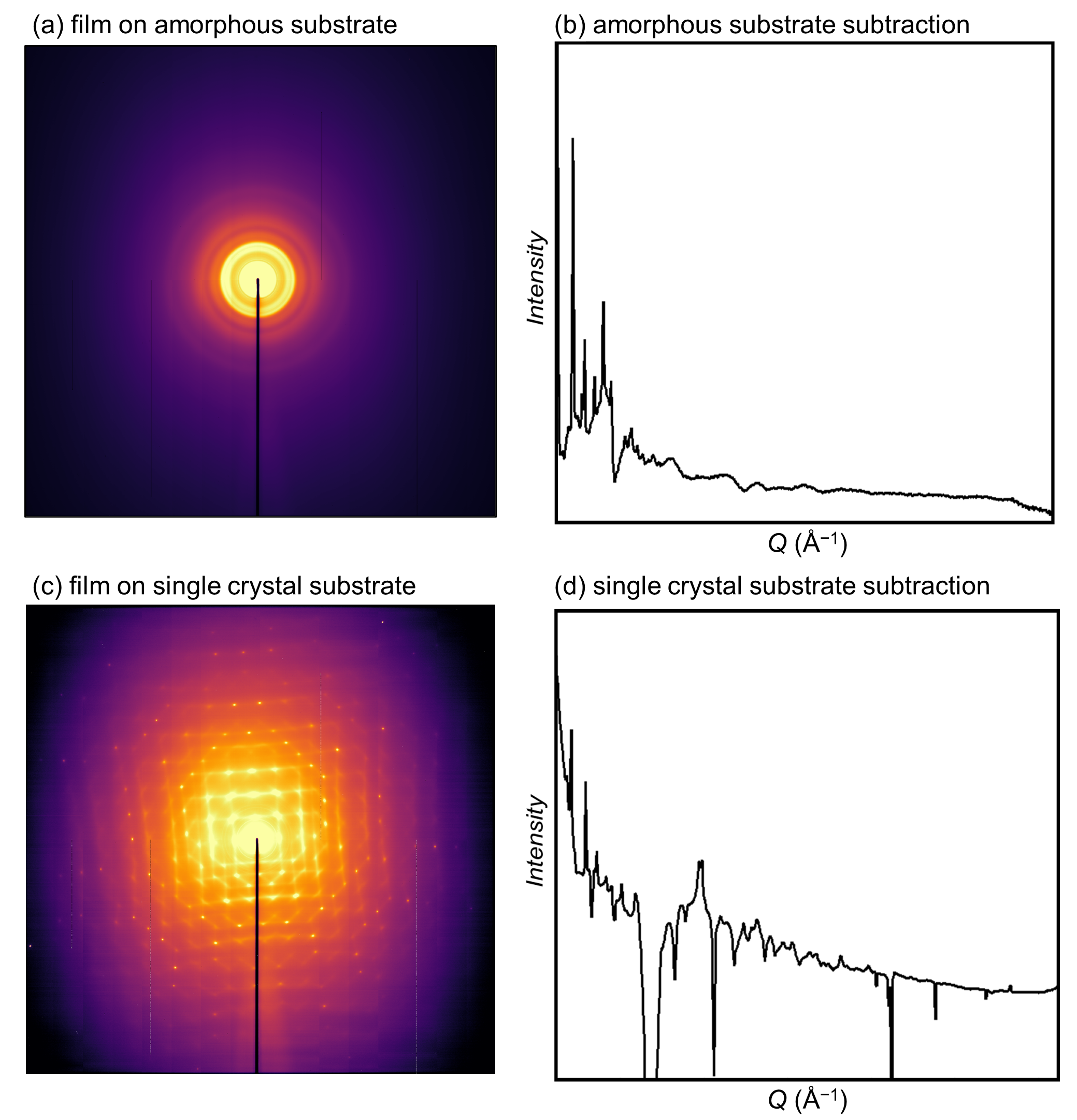}
\caption{2D synchrotron detector images of 300\,nm thick Ge$_2$Sb$_2$Te$_5$ X-ray total scattering signals measured at 0.1667\,\AA\/ (74\,keV) deposited on (a) 500\,$\mu$m thick amorphous SiO$_2$ substrate and (c) 500\,$\mu$ thick single crystal Si (100) substrate. The substrate subtraction method was applied to obtain the $I$($Q$) data for (b) the amorphous substrate sample and (d) the single crystal substrate sample.}
  \label{fig:ampvssc}
\end{figure}

The smooth, directionally-independent scattering signal of amorphous substrates is advantageous for thin film PDF measurements, which makes the scaling and subtraction of substrate scattering relatively straight forward (Fig.\,\ref{fig:ampvssc}a, b).\cite{jensen_demonstration_2015, Shyam2016} In contrast, scattering data for a similar thin film deposited on a (100)-oriented single crystal Si substrate has high intensity Si reflections that appear as spots in the two-dimensional (2D) image (Fig.\,\ref{fig:ampvssc}c). With a single crystal substrate, subtraction of the scattering pattern from a blank substrate is ineffective, resulting in both residual signal and oversubtraction of the high intensity reflections. The resulting 1D scattered intensity, $I$, as a function of momentum transfer, $Q$, [$I$($Q$)]has contributions from the substrate in some places, and unphysical negative scattered intensity elsewhere (Fig.\,\ref{fig:ampvssc}d). This is due to differences in the relative intensities of diffuse and Bragg scattering of the substrate with and without a thin film.\cite{kabsch_evaluation_1988} The resulting subtracted signal would thus be inappropriate for generating PDF data. Therefore, an alternative to the subtraction process is needed to elucidate accurate structure$-$property relationships for thin film samples on single crystal substrates, which alter their structure and properties.

To more effectively separate the X-ray total scattering signal of the thin film from that of its substrate, we use NMF to differentiate between the directionally-dependent and directionally-independent features. NMF and other machine learning algorithms have been previously applied in materials measurement science to classify and analyze data.\cite{holm2020overview,liu2020data,wang2017x} Of particular relevance to this work, unsupervised machine learning algorithms excel in treating unlabeled input datasets and, in turn, generating organized or sorted datasets of interest.\cite{Tetefr_2022,wei_machine_2019} Among these algorithms, non-negative matrix factorization (NMF) is a dimensionality reduction algorithm that accepts inputs from a non-negative matrix (composed solely of positive values) and produces feature and weight matrices that reconstruct the original data.\cite{Wang_NMF_2013, Stanev_NMF_2018} NMF is particularly useful for X-ray scattering data, which inherently only have positive values, especially for \textit{in situ} experiments that generate 20 or more patterns.\cite{elton1966x, maffettone_constrained_2021, Long2009, Thatcher2022} 

We report a novel data processing method for X-ray total scattering data from thin films on single crystal substrates, IsoDAT2D. For an individual sample at ambient conditions, X-ray total scattering data are collected as a single 2D image, which is azimuthally integrated to a single, one-dimensional (1D) $I$($Q$) dataset.\cite{Filik2017, Prescher2015, Ashiotis2015} A single dataset is incompatible with NMF, which identifies features across a series of individual datasets that occur with different weightings.\cite{Dou2023} The developed method uses a novel integration process that not only provides the data necessary for NMF, but also takes advantage of the differences of scattering features from the film and substrate. Using these input datasets for NMF results in many outputted `feature' components. We use unsupervised machine learning clustering algorithms, specifically hierarchical agglomerative clustering (HAC), to group similar outputted components based on their features, which are expected to reflect their origin (\textit{i.e.}, film \textit{vs.} substrate).\cite{Murtagh2011, Iwasaki2017} After clustering, components from one or more clusters are averaged and smoothed to thin film X-ray total scattering data of sufficiently quality for PDF analysis. 

We developed and validated IsoDAT2D using synthetic data generated by our novel thin film X-ray scattering data creation program, SimDAT2D. We demonstrate IsoDAT2D on experimental X-ray total scattering thin film data, successfully isolating the isotropic scattering signal of a thin film on a single crystal substrate. Using the isolated signal, PDF data quality is significantly improved relative to substrate subtraction using a single crystal substrate. Further, PDF data from IsoDAT2D are of comparable quality to PDF data from amorphous substrate subtraction.

\section{Developed Programs}

\subsection{SimDAT2D: Generating Synthetic 2D Scattering Data}

\begin{figure}
\graphicspath{ {./Figures/} }
\includegraphics[width=4.5in]{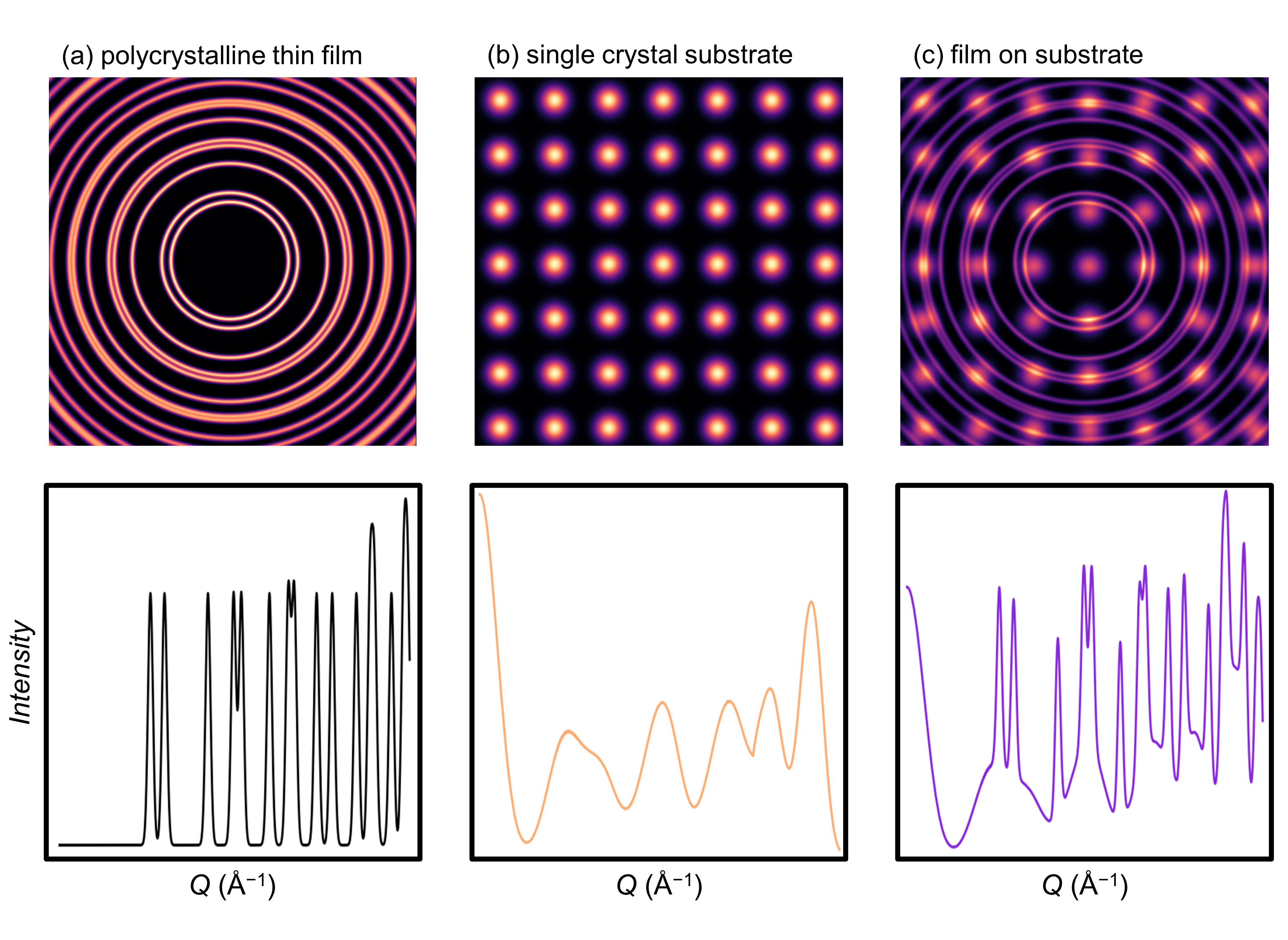}
  \caption{Synthetic scattering data generated with SimDAT2D. (a) Simulated isotropic scattering of LaB$_6$ generated using a 500\,mm sample-to-detector distance and an X-ray wavelength of 0.5\,\AA. (b) Simulated anisotropic scattering signal for a cubic single crystal. (c) A linear combination of (a) isotropic and (b) anisotropic signals representing total scattering data from a thin film on a single crystal substrate.}
  \label{fig:tfdc}
  \end{figure}

 A key aspect of SimDAT2D is its use to generate synthetic X-ray scattering data from two or more distinct scattering contributions (\textit{e.g.}, isotropic, diffuse, anisotropic; Fig.\,\ref{fig:tfdc}). Each scattering contribution is generated individually and combined as a weighted linear combination into a single 2D detector image. This allows for the relative intensities to be varied and for each signal to be independently known. We created SimDAT2D to enable the generation of synthetic 2D X-ray scattering data and masks, as well as using those to perform azimuthal and rotational integrations of 2D data using the PyFAI Python library.\cite{Ashiotis2015}

Isotropic signals consisting of radially symmetric rings at discrete positions over a 2D detector image are created using PyFAI's calibrant materials (30 unique options), taking into account user-defined wavelength, sample-to-detector distance, and detector type (\textit{e.g.}, PerkinElmer Silicon)(Fig.\,\ref{fig:tfdc}a). Anisotropic scattering signals, with diffraction spots at discrete positions on the 2D image, \textit{e.g.,} the scattering of a single crystal substrate, can be created manually as a grid of spots (Fig.\,\ref{fig:tfdc}b); for this, the user specifies the number of spots per line, the distance between spots, the diameter of spots (pixel width), and the Gaussian distribution of spot intensity. Alternatively, scattering patterns of any kind can be incorporated, for example, drawing upon existing tools to simulate patterns, such as SingleCrystalMaker\textsuperscript{\textregistered}, which calculates single crystal diffraction patterns.\cite{palmer2015visualization} Likewise, experimentally measured 2D detector images can be used.

Individual images are combined in a linear combination in with user defined relative weightings of each contribution (Fig.\,\ref{fig:tfdc}c). For example, to create an image that resembles a thin film on a single crystal substrate, increasing the relative weighting of the generated substrate signal results in intensity differences that qualitatively resemble experimental data. The images to be combined must have the same pixel dimensions; specifically, the program uses a 2048\,$\times$\,2048 pixel grid, which is standard for many X-ray total scattering detectors. 

In addition to combining multiple scattering patterns, SimDAT2D scattering patterns can incorporate noise of various levels, allowing for synthetic data that more closely resembles experimental data. Noise is generated as a 2D map as a randomized normal distribution of coefficients by which synthetic data are multiplied (Fig. S6).

\subsection{IsoDAT2D: Identifying Isotropic Scattering Signals}

\begin{figure}
\graphicspath{ {./Figures/} }
\includegraphics[width=6 in]{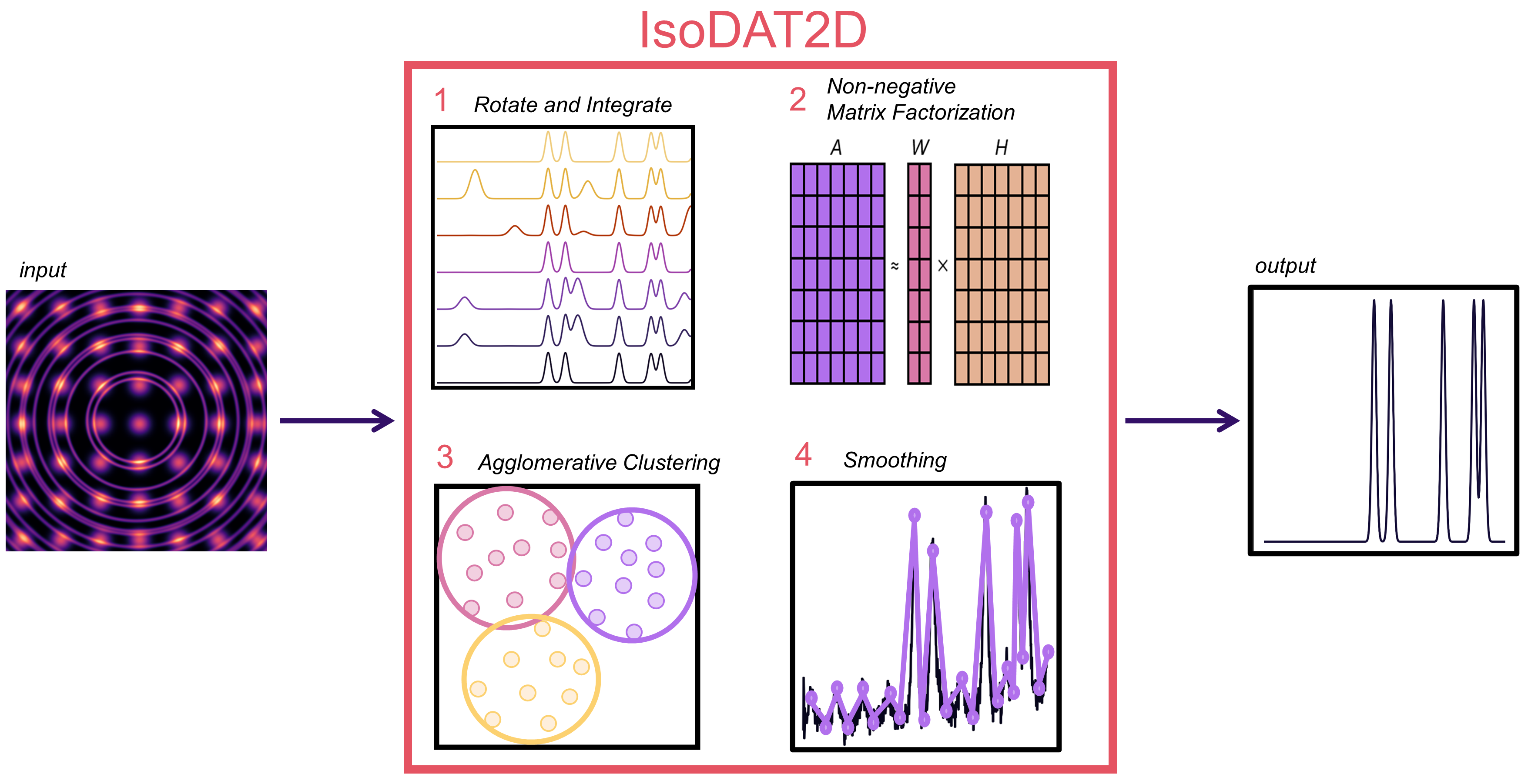}
\caption{IsoDAT2D processes a 2D total scattering detector image input in four steps: (1) rotational integration generates line integrations that are inputs for (2) NMF. The components identified by (2) NMF are (3) clustered based on their similarities using HAC. The clusters of interest from (3) are then (4) averaged and smoothed using a Savitzky-Golay filter, outputting a single 1D X-ray scattering dataset.}
  \label{fig:IsoDAT2D}
\end{figure}

We developed IsoDAT2D to identify isotropic scattering data from 2D detector images comprising isotropic and anisotropic contributions. From this, thin film PDF data can be acquired. The program processes thin film X-ray total scattering data using NMF and HAC (Fig.\,\ref{fig:IsoDAT2D}). This combination of machine learning algorithms eliminates the need for substrate subtraction, which is effective for films on amorphous substrates but not single crystal substrates.\cite{Dippel:ro5016, jensen_demonstration_2015}

The first step of IsoDAT2D is a `rotation and integration' (Fig.\,\ref{fig:IsoDAT2D}-1), which generates tens to thousands of 1D datasets from a single 2D image, as is needed for machine learning algorithms. These 1D datasets are inputs for NMF, (Fig.\,\ref{fig:IsoDAT2D}-2), which identifies recurring components over the series of datasets. Next, HAC is used to sort the components generated by NMF (Fig.\,\ref{fig:IsoDAT2D}-3), such that one or more of the resulting clusters contains components associated with the scattering pattern of the thin film. Of those clusters, one or more comprise components that resemble isotropic scattering (Fig.\,S4). The components in the cluster(s) are then averaged to a single $I$($Q$) X-ray total scattering dataset. The dataset is smoothed to reduce noise introduced through the data processing (Fig.\,\ref{fig:IsoDAT2D}-4), ultimately returning an output of X-ray total scattering data from which PDF data can be derived. These steps and key parameters are further discussed following, with example inputs and outputs based on data generated using SimDAT2D. Details on data generation for the demonstrative example are described in Supplementary Information (Section S1.1).

\subsubsection{Rotation and Integration}
\begin{figure}
\graphicspath{ {./Figures/} }
\includegraphics[width=6in]{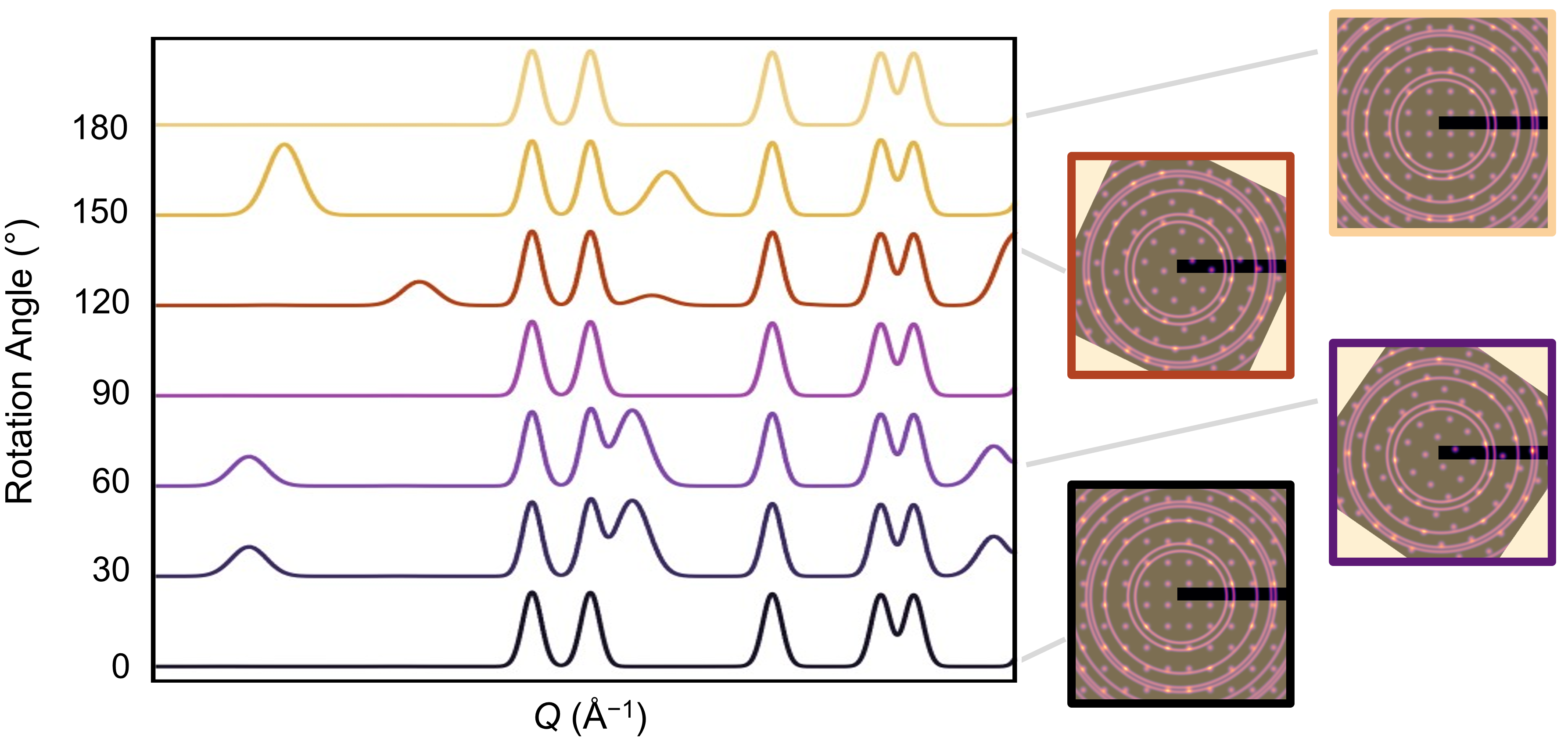}
\caption{A waterfall plot of line integrations from a single 2D detector image rotated at 30$^{\circ}$ intervals. The isotropic rings in 2D appear in the same $Q$ locations in 1D data for all rotations, while the anisotropic single crystal pattern change with rotation angle.}
  \label{fig:rotandint}
\end{figure}

2D X-ray total scattering data are typically azimuthally integrated into a single 1D $I$($Q$) dataset. In the current data processing workflow, multiple narrow integrations are used instead. These allow us to exploit the inherent difference between radially-dependent (anisotropic) patterns of Bragg spots from a single crystal substrate and radially-symmetric (isotropic) scattering from an amorphous or polycrystalline thin film (Fig.\,\ref{fig:rotandint}). These many discrete integrations also enable the use of machine learning methods for data processing, which require multiple input datasets.\cite{RamezanTrainingSetSize, zhang2018strategy}

Bragg diffraction from single crystals occurs at discrete points in space,\cite{Kabsch1988} resulting in high-intensity spots in patterns that reflect the symmetry of the crystal structure.\cite{warren1990x} In addition, high-quality single crystal substrates, such as single crystal Si, have phonon-phonon interactions that contribute weak--but detectable--diffuse scattering (Fig.\,\ref{fig:ampvssc}c).\cite{Shestakov2018, Lonsdale_1941} Powder samples can be approximated as a collection of small single crystals with random orientations relative to one another. This gives discrete points across all positions of scattering cones, creating isotropic rings in detector images.\cite{Harris2001, Sivia2011} Likewise, untextured thin films (\textit{e.g.}, polycrystalline, nanocrystalline, amorphous) have radially-symmetric scattering. 

X-ray total scattering data for PDF analysis are typically acquired at synchrotron facilities in transmission geometries over a wide $Q$ range using 2D detectors.\cite{Jrgensen2018,Nakamura2020, DiazLopez2020} Our rotational integration uses these 2D scattering images with other standard integration inputs, such as detector calibration parameters (\textit{e.g.}, PONI in PyFAI), sample-to-detector distance, and X-ray wavelength.\cite{Ashiotis2015} 

For standard X-ray total scattering measurements, masks are used to exclude features from the integration, such as signals from a beam stop and dead detector pixels, covering no more than $\approx$20\% of the image area. In contrast, we use masks to discretize the detector area to generate many 1D linescans. For this, masks leave only a small fraction of the detector area exposed, $\approx$1\%; thus only small image areas are integrated at a time (Fig.\,\ref{fig:rotandint}, right). After an integration, the image is rotated by a user-defined angle and an integration is taken over the new exposed area. This is repeated until the whole 2D area is integrated. 

Figure\,\ref{fig:rotandint} shows an example of line integrations taken at 30$^{\circ}$ rotational intervals. In the resulting integrated data, a set of peaks occurs at the same position and intensity in each dataset; this is the isotropic signal from the thin film. Other features change position and intensity between datasets; these are from the anisotropic Bragg spots from the single crystal substrate. 

The shape and size of the integrated area can be modified by selecting the mask(s) and the number of integrations can be varied by the degree of rotation (see Supplementary Information, Section S1.2). For the NMF input, having a sufficient number of datasets is essential for identifying the isotropic scattering signal, which is enabled by the redundancy of those features across integrated input 1D datasets. 

To generate larger data series with various contributions of isotropic and anisotropic signals, 1D datasets from a combination of mask geometries and rotational angles can also be used. For example, masks with various numbers of sectors and differences in their widths and offsets can be used. We investigated this `multi-mask' approach in application to SimDAT2D-generated synthetic data, the results of which are described in the Supplementary Information (Section S1.3).

The frequent occurrence of features enhances the likelihood of the desired signal being successfully identified by NMF.\cite{Wang2013} To balance redundancy and computational expense we find that a minimum of 360 integrated datasets per 2D image should be used (Fig. S3). However, with access to computational resources, the use of more input datasets improves the efficacy of signal identification describes in the Supplementary Information (Section S1.3).

\subsubsection{Non-Negative Matrix Factorization}

NMF was originally developed to reduce the amount of data required to describe a system or phenomenon. Its ability to do this can be used to identify meaningful components in a series of data. The `non-negative' in NMF reflects that inputted data, identified components, and the weights of those components contain only positive values.\cite{gillis2014and}

The mathematical process behind NMF is designed to decrease the Euclidean distances between separate non-negative matrices. To achieve this, a Frobenius norm loss function, $\left \| A - WH \right \|_{loss}$, is implemented as a metric for the magnitude of the difference between the inputted and feature matrices. The features matrix, $H$, contains components that, when weighted according to matrix $W$, reproduce the original inputted matrix, $A$.\cite{lee2000algorithms} In the context of this work, the $H$ matrix contains components that correspond to scattering from specific sample elements, \textit{i.e.}, the single crystal substrate and the thin film.

\begin{figure}
\graphicspath{ {./Figures/} }
\includegraphics[width=2.5in]{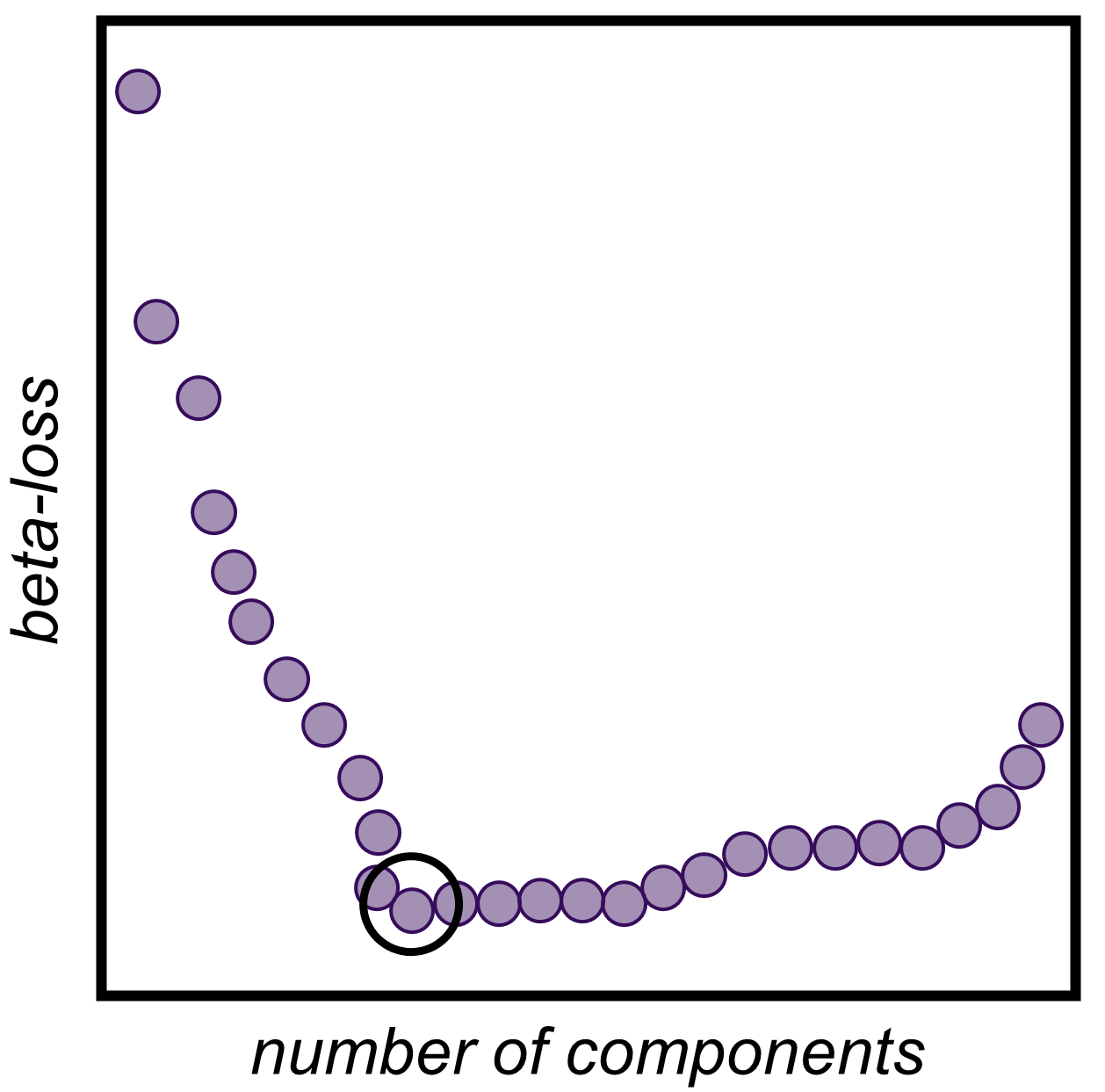}
  \caption{An elbow plot illustrating how the beta-loss function varies with the number of components for the NMF algorithm. The black circle highlights the lowest beta-loss point, representing the position where the number of components and the beta-divergence are at a minimum, indicating the optimal number of components for feature identification.}
  \label{fig:elbow}
\end{figure}

In our use of NMF, we employ a process that minimizes the Euclidean distance using beta-divergence as the metric to iterate through different numbers of output components in the feature matrix. \cite{Olaya_beta_divergence} To illustrate, when the original input matrix contains 360 datasets, our application runs NMF with the number of component values from 2 to 360, and then selects the smallest number of components that produce the lowest beta divergence (Fig.\,\ref{fig:elbow}).

We implement Scikit-learn's NMF algorithm, which is initialized with randomized parameter variables, including initializer, solver, beta-loss, and tolerance.\cite{scikit-learn} In our implementation of NMF, the series of 1D $I$($Q$) integrations generated from rotation and integration are used as the original matrix, $A$. We use the data as integrated, without normalization, which enables us to later sort the data based on variance that would be lost with normalization.\cite{Maisog2021}

The initializer defines the method selected to run the program, the solver determines which numerical optimizer is used, beta-loss specifies the beta-divergence minimization function, and the tolerance indicates the proximity to a specific value of beta-divergence that must be reached to end the process.\cite{Flores2022} The described NMF process is repeated for a user-defined number of iterations and returns the feature matrix, $H$, with the minimum beta-divergence. The number of components at which the beta-divergence is minimized appears as a discontinuity in the beta-loss as a function of the number of components (Fig.\,\ref{fig:elbow}). This `elbow' indicates when the amount of components required to recreate the dataset has been identified. 

Depending on the nature of the inputted data, the identified feature matrix may contain thousands of components. A challenge of NMF is appropriately sorting the identified components. To manage this, we implement data science tools to help identify the component(s) that correspond to the thin film X-ray total scattering signal.

\subsubsection{Hierarchical Agglomerative Clustering}

\begin{figure}
\graphicspath{ {./Figures/} }
\includegraphics[width=3.25in]{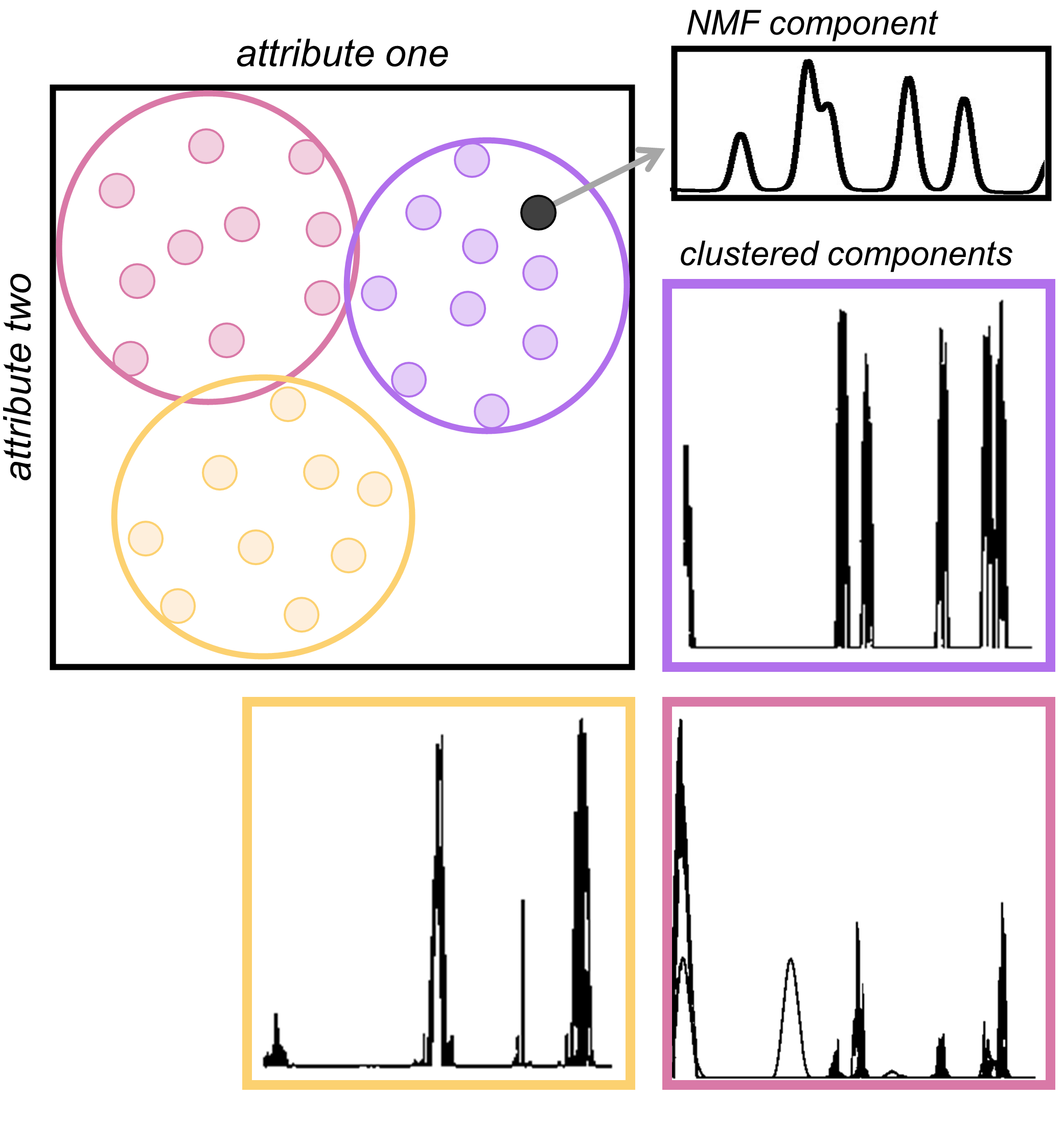}

  \caption{Schematic representation of HAC in two dimensions. Individual NMF output components (filled circles) are sorted into clusters (open circles) based on variance. Examples of all of the datasets within three clusters have similar features (boxes).
  }
  \label{fig:hac}
\end{figure}

We use HAC to efficiently sort components identified by NMF (Fig.\,\ref{fig:hac}). The process groups NMF components based on similarities, creating `clusters' of like components. As a `bottom-up' approach, agglomerative clustering initiates many small clusters that are merged to create larger ones, unlike a `top-down' approach, in which a single large cluster is progressively divided into smaller structures.\cite{sasirekha2013agglomerative,chowdhury2022assessment} Using agglomerative clustering, we are able to assign the number of initial clusters into which outputted NMF components must converge.

For the implementation of agglomerative clustering in IsoDAT2D, the number of clusters is user-defined and informed by the number of scattering phenomena expected to contribute to the measured total scattering signal. The algorithm should produce separate clusters associated with each type of scattering, for example, at least two clusters for single crystal substrates, which have diffuse scattering and Bragg spots.

We use variance as the cluster sorting parameter, which supports the assignment of components to appropriate clusters in our unnormalized data. This leverages high variance of scattering intensity from features associated with the thin film and the substrate, which results from their relative volumes. Specifically, our clustering approach employs Ward's method, a widely-used linkage criterion that minimizes the Euclidean distance between points in the parameter space.\cite{Murtagh2011,szekely2005hierarchical} Ward's method is particularly effective for our data as it efficiently accounts for the variations in scattering intensities to ensure robust clustering. This allows us to identify and separate signals of interest from the NMF output matrix components, enhancing the quality of the resulting data and providing valuable insights into the scattering phenomena.

After HAC, the 1D datasets in each cluster are plotted and visually evaluated by the user to assess if the data in the cluster resemble isotropic scattering. We found there to typically be multiple clusters with these contributions (Fig. S3 \& 4). Further, the identified components within the clusters contain moderate noise from previous data processing steps. To reduce processing noise and have a singular 1D dataset, an approach has been created to average and smooth the identified components. 

\subsubsection{Averaging and Smoothing}

In typical total scattering data processing, azimuthally integrating 2D images improves counting statistics and the resulting data quality. Here, the individual line integrations do not have this benefit, but statistics are improved by merging clustered components, which, despite being in one cluster, are not identical. This averaging is reminiscent of the angular averaging inherent to the higher data quality achieved with azimuthal integration. \cite{Brgemann2004} 

Components of the selected cluster(s) are normalized to uniform scaling and then averaged to a single 1D dataset. However, the narrow integrations and NMF results in more noise than an azimuthal integration. To address this, we implement a Savitzky-Golay filter from the SciPy library, which uses a polynomial fit to smooth high-variance regions in the data.\cite{press1990savitzky,2020SciPy-NMeth} This reduces noise while preserving key features of the data. Additional considerations for working with noise are described in the Supplementary Information (Section S1.4). 

\section{Experiments and Results}
\label{sec:results}

We validated the capabilities of IsoDAT2D to separate signals of interest using synthetic data generated by our SimDAT2D program. Using these synthetic data, in which the `answer' data could be simulated directly and compared to the identified signal, we assessed the effects of parameters such as masking, noise, and integration resolution (Supplementary Information, Section S1).\cite{John2022} Experiments using synthetic data, described in the Supplementary Information (Section S1.1) were used to qualitatively evaluate IsoDAT2D's performance and inform parameters to minimize noise and optimize isotropic signal identification.

Building on the insights gained from synthetic data experiments, we applied IsoDAT2D to experimental X-ray total scattering data collected from a 300\,nm thick polycrystalline film of Ge$_2$Sb$_2$Te$_5$ (GST) on a 500\,$\mu$m single crystal Si substrate (synthesis and data acquisition information provided in Supplementary Information, Section S2.1). We qualitatively compare the $I$($Q$) and $G$($r$) data for this film processed using IsoDAT2D and substrate subtraction. We also compare data from a film of the same composition and thickness on an amorphous SiO$_2$ substrate processed using substrate subtraction.

\subsection{Substrate Subtraction}

\begin{figure}
\graphicspath{ {./Figures/} }
\includegraphics[width=3.25in]{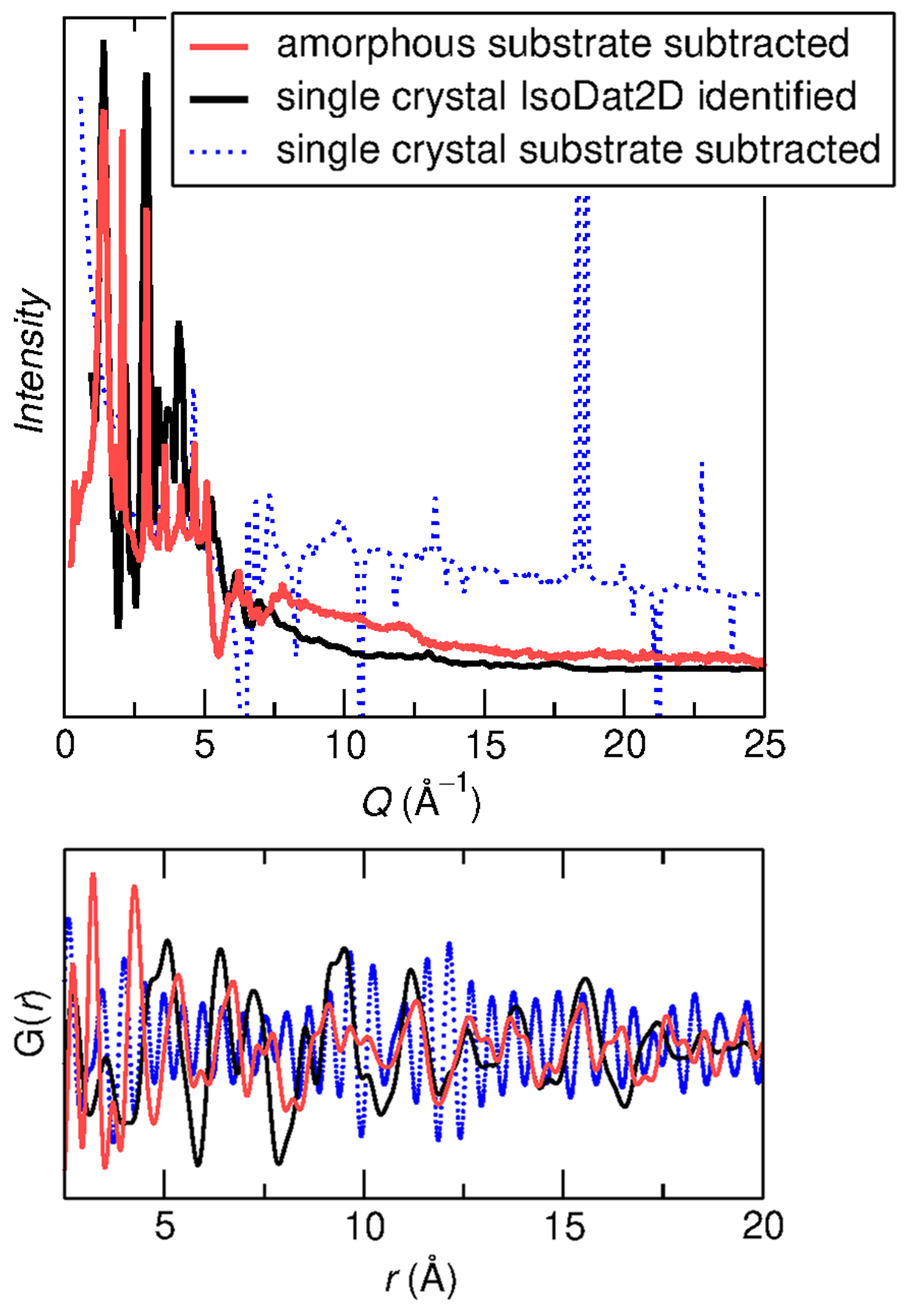}
\caption{Substrate-subtracted and IsoDAT2D-identified signals from crystalline GST thin films deposited on amorphous and single crystal substrates. The top panel illustrates the $I$($Q$) X-ray total scattering data and the bottom panel shows the corresponding PDFs, $G$($r$), derived from the X-ray total scattering data.}
    \label{fig:gst_xrd_gr}
\end{figure}

Following previous methods, the X-ray total scattering data for GST on SiO$_2$ and Si substrates were isolated by integrating 2D data collected of the thin film samples and of the substrates without films, then scaling and subtracting the latter from the former.\cite{jensen_demonstration_2015} 
For this process, 2D detector images were azimuthally integrated to 1D linescans with identical masks using PyFAI\cite{Ashiotis2015}, resulting in 1D linescans. The $I$($Q$) of the substrates were scaled to maximize their intensity without exceeding the intensity of the corresponding thin film sample data. The scaled data from the substrate were then subtracted from the film-on-substrate data, leaving a signal comprising data from only the thin film (Fig.\,S9, Fig.\,\ref{fig:gst_xrd_gr}). Using PDFgetX3,\cite{Juhs2013} the substrate-subtracted 1D linescan was Fourier transformed to PDF data with $Q_{max}$ = 14\,\AA$^{-1}$ (Fig.\,\ref{fig:gst_xrd_gr}).

\subsection{IsoDAT2D Component Identification}

X-ray total scattering data from the GST thin film on a single crystal Si substrate were also processed using IsoDAT2D. A single 2D detector image was sequentially rotated and integrated using several combinations of parameters (\textit{e.g.}, mask geometry, number of datasets, resolution, etc.). The resulting 1800 datasets with 2500 data points resulting from the use of five separate masks balanced the clarity of signal features and noise effects. 

For rotation and integration, five iterations of single and multi-sector masks were used with 1, 2, 3, 4, and 5 slices of 0.5\,px width rotated every 1\(^{\circ}\) to produce 1800 1D datasets (Supplementary Information, Section S1.3). Similar to experiments with synthetic data, using these experimental data we found that this number of datasets balanced computational efficiency and data quality, especially the signal-to-noise ratio (Fig.\,S5).

We selected a data point resolution for integration of 2500 points, which balanced signal identification and noise effects (Supplementary Information, Section S1.4). With higher data point resolutions, the program's output had significant noise. Using lower data point resolutions, the program identified signal features that were unphysical, especially peak widths and isolated signals that were not suitable for PDF analysis.

From the generated 1D datasets, NMF components that resemble the GST signal of interest were identified in several clusters. Data from these clusters were averaged and smoothed to a single $I$($Q$) dataset. These data required some further processing for the intensity to approach zero with increasing $Q$ due to processing effects. For this, the value of the lowest intensity point was subtracted from the data. Subsequently, the dataset was multiplied by a Gaussian function to mitigate unphysical high-$Q$ features, resulting in $I$($Q$) data suitable for PDF analysis. The $I$($Q$) data were transformed to $G$($r$) with PDFgetX3 using a $Q_{max}$ = 14\,\AA$^{-1}$.\cite{Juhs2013}

\subsection{Comparison of Data from Each Processing Method}

Identified X-ray total scattering data from amorphous substrate subtraction and IsoDAT2D are qualitatively similar (Fig.\,\ref{fig:gst_xrd_gr}). In contrast, single crystal Si substrate subtraction data has unphysical features due to the oversubtraction of high intensity single crystal Bragg spots. Data for amorphous substrate subtraction are aligned with our expectations of X-ray total scattering data, especially the decrease of intensity of features with increasing $Q$. The IsoDAT2D-identified $I$($Q$) signal is similar in its overall variation of intensity with $Q$, affirming the efficacy of the processing method. The positions of peaks in these two $I$($Q$) signals are also consistent, with the primary difference between the identified signals being peak geometry, especially peak widths and the relative intensities of reflections in each pattern. Overall, this comparison highlights that the $I$($Q$) data from IsoDAT2D are comparable with those from the previous method used for thin film PDF, highlighting the promise of accessing atomic structure information from thin films deposited on single crystal substrates using X-ray total scattering. \cite{jensen_demonstration_2015}

\begin{figure}
    \centering
    \includegraphics[width=6in]{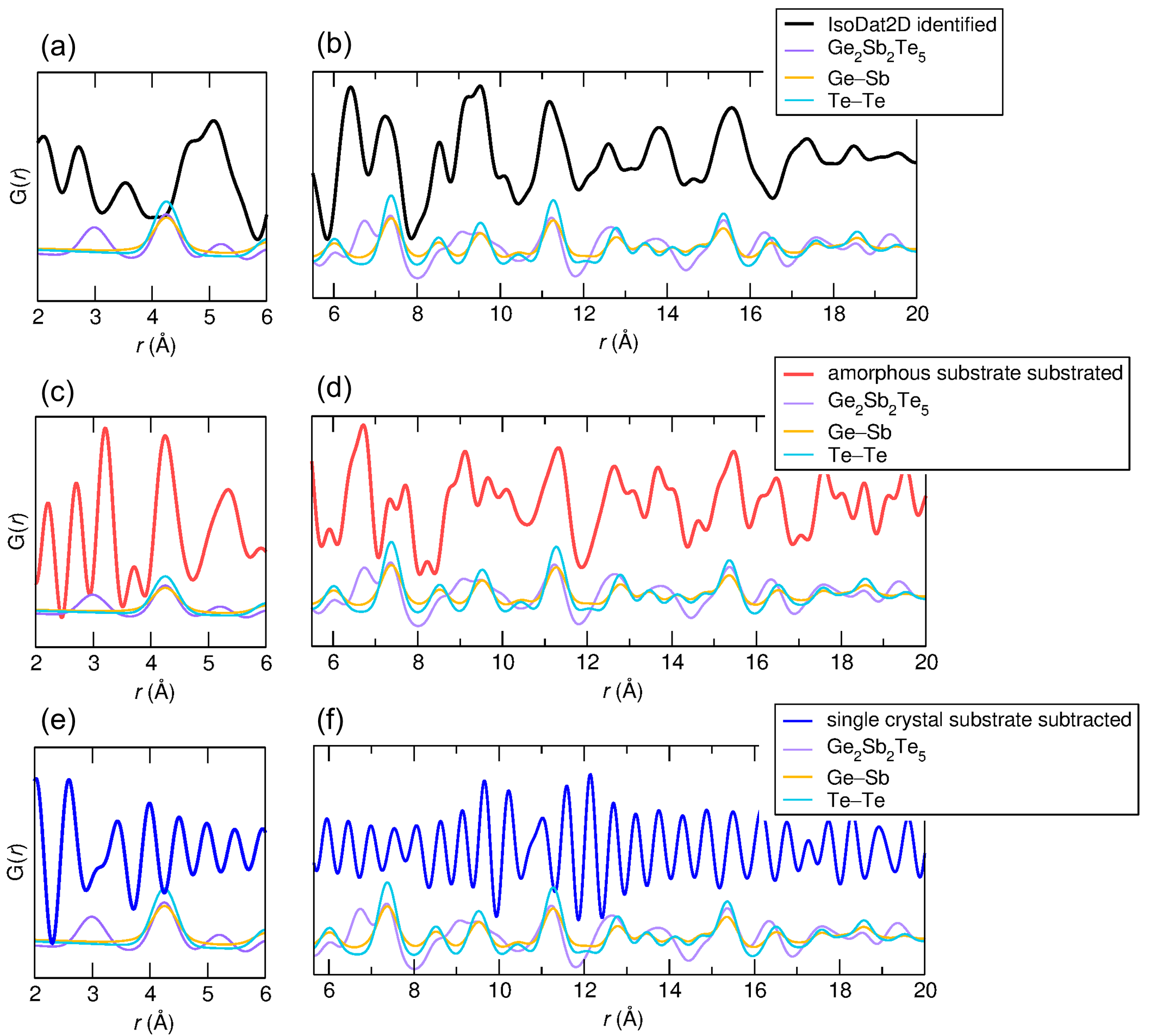}
    \caption{Comparison of local and mid-range PDF of thin film GST to calculated PDF and Ge$-$Sb and Te$-$Te partials.\cite{shamoto2006local} Local and mid-range structure ranges of thin film PDF data from GST on (a,b) an amorphous substrate processed by substrate subtraction, (c,d) a single crystal substrate processed with IsoDAT2D, and (e,f) a single crystal substrate processed by substrate subtraction.}
    \label{fig:partials_pdf}
\end{figure}

The $I$($Q$) data from each processing method were Fourier transformed to generate PDF data (Fig.\,\ref{fig:gst_xrd_gr} bottom) using PDFGetX3.\cite{Juhs2013} The PDF, $G$($r$), were compared to one another, as well as to calculated PDFs and Te$-$Te and Ge$-$Sb partial PDFs for GST. Calculated data were based on a GST model from neutron PDF analysis\cite{shamoto2006local} and were calculated using PDFgui (Fig.\,\ref{fig:partials_pdf}) \cite{farrow2007pdffit2}.

Similarities and differences between the $I$($Q$) signals are reflected in the resulting PDF data. For amorphous substrate subtraction and IsoDAT2D, PDF data are qualitatively similar, with correlations consistent with our expectations for a polycrystalline material. In contrast, PDF data from the single crystal substrate subtraction show termination ripples and high-frequency noise, making it entirely unsuitable for deriving even qualitative information about atomic structure. 

In the low $r$ range, PDF data from amorphous substrate subtraction and IsoDAT2D processes miss the first correlation centered at 3\,\AA\,(Fig.\,\ref{fig:partials_pdf}a \& c). For the second and third correlations, the amorphous substrate-subtracted data match calculated correlations between 4\,\AA\, and 6\,\AA \,(Fig.\,\ref{fig:partials_pdf}c). Local structure correlations in IsoDAT2D-identified data are less consistent with calculated data (Fig. \ref{fig:partials_pdf}a), exhibiting a single broad correlation at $\approx$5\,\AA\/ rather than the two correlations near 4.2\,\AA\/ and 5.2\,\AA. At higher $r$ ranges, representing mid-range and average structure features, both the amorphous substrate-subtracted and IsoDAT2D-identified signals exhibit strong similarities with the calculated signal, accounting for most of the correlations (Fig.\,\ref{fig:partials_pdf} b, d). Overall, the PDF data from the IsoDAT2D process is of similar quality to the data from amorphous substrate subtraction.

\section{Discussion}

While there are established methods for identifying the scattered signal from thin films on amorphous substrates, the same approach cannot be applied to films on single crystal substrates (Fig.\,S9). To address this mismatch between measurement needs and capabilities, we developed IsoDAT2D to enable the measurement of atomic structures for amorphous and nanocrystalline thin films on single crystal substrates, a class of materials that have not been previously measurable with transmission geometry X-ray total scattering. IsoDAT2D effectively isolates the total scattering signal from a thin film on a single crystal substrate, from which PDF data for the film can be generated. With this approach, we enable PDF analysis of polycrystalline, nanocrystalline, and amorphous thin films on single crystal substrates without compromising $Q_{max}$, as in grazing incidence experiments.\cite{Dippel:ro5016}

The method we report leverages inherent differences in the 2D symmetry of the scattering patterns of the film and substrate, using hundreds to thousands of line integrations rather than a single azimuthal integration over the 2D detector image. The resulting 1D integrations are inputs for NMF, from which identified components are sorted using HAC, ultimately returning an output that is the X-ray total scattering of the isotropic component (\textit{i.e.}, the thin film). 

Using synthetic data, we found that increasing the number and variety of features in 1D integrations by using a variety of mask geometries and orientations increases the likelihood of identifying the signal of interest and decreases the noise associated with these narrow integrations. However, increasing the amount of input data also increases computational expense, especially for the currently implemented NMF algorithm, which requires both significant memory and computational power.

In addition to high-fidelity identification from synthetic data, IsoDAT2D identified the thin film scattering signal from experimental data. The thin film scattering signal returned by the algorithm was consistent with that identified using conventional substrate subtraction methods from an analogous thin film on an amorphous substrate. 

Thin film PDF data are challenging to collect, and we find that both the amorphous substrate-subtracted and IsoDAT2D-identified scattering signals deviate from expectations in the local structure. However, their mid-range structures were consistent with one another and simulated PDF data of the published structure.\cite{shamoto2006local} While the amorphous substrate subtraction and IsoDAT2D approaches effectively capture mid-range correlations, the local structure features from both methods differ from the calculated PDF. Since both methods have discrepancies between measured and calculated local structure, this could be a common limitation, or at least challenge, of applying PDF to thin film samples. Similarly, substrate subtraction and IsoDAT2D are so far limited to relatively thick films, with successful demonstration previously and here on films $\geq$ 300\,thick.\cite{jensen_demonstration_2015} However, having validated the capabilities of this machine learning data processing approach, improvements can be made to further push the boundaries of the types of samples to which PDF analysis can be applied, such as thinner films.

Due to the similar strengths and shortcomings of both processing methods, we conclude that IsoDAT2D enables similar quality PDF data as the substrate subtraction method, but can uniquely be applied to films on single crystal substrates. In the illustrative example shown here, to probe the crystalline structure of a phase change material, differences in structure and, accordingly, properties are not expected to depend on the substrate. However, IsoDAT2D enables PDF data from thin film samples for which structure and function are directly affected by the substrate. \cite{saha2002effects, ghosh2004effect, li2002effect}

\section{Conclusions}

We developed IsoDAT2D to identify thin film X-ray total scattering signals from signal crystal anisotropic signals. Combining a novel approach for integrating 2D scattering images with NMF and HAC machine learning algorithms, IsoDAT2D effectively identifies isotropic scattering signals.

To validate the algorithm's performance, we created SimDAT2D, which generates synthetic 2D images representative of scattering from thin films on single crystal substrates, allowing for the comparison of algorithm-identified signals with a known synthetic signal. Further, SimDAT2D is a valuable tool for exploring the effects of various parameters on algorithm performance, offering insights into IsoDAT2D's capabilities and informing data processing approaches specific to the nature of the film and substrate.

Beyond thin film characterization, IsoDAT2D and similar approaches may be applicable to other materials systems and data in which there are constant and dynamic signals of interest, \textit{i.e.}, from \textit{in situ} and \textit{operando} scattering experiments. By advancing our understanding of structure$-$property relationships in application-relevant thin film materials, these tools pave the way for fundamental and technological advancements.

\section{Conflicts of Interest}
None to report.

\section{Acknowledgments}
The authors acknowledge the University of Florida Research Computing for providing computational resources and support that have contributed to the research results reported in this publication. This research used the Pair Distribution Function Beamline (PDF) of the National Synchrotron Light Source II, a U.S. Department of Energy (DOE) Office of Science User Facility operated for the DOE Office of Science by Brookhaven National Laboratory under Contract No. DE-SC0012704. The authors would also like to acknowledge Dr. David Adams at Sandia National Laboratory for providing thin film samples. 

\bibliography{references.bib}



\section*{Supplementary material}

# Distinguishing Isotropic and Anisotropic Signals for X-ray Scattering Data through Machine Learning

Danielle N. Alverson,[†] Daniel Olds,[‡] and Megan M. Butala[*,†]

†*Department of Materials Science and Engineering, University of Florida, Gainesville, FL 32611, USA*

‡*National Synchrotron Light Source II, Brookhaven National Laboratory, Upton, NY 11973, USA*

E-mail: mbutala@ufl.edu

# Supplementary Information

## S1 Synthetic Thin Film Data Parameterization

### S1.1 Methods

Synthetic 2D scattering images representative of a thin film on a single crystal substrate were created using SimDAT2D. Isotropic rings (polycrystalline film) and an array of spots (single crystal substrate) were generated separately and then combined in a single 2D pattern. The isotropic scattering pattern (rings) and the anisotropic pattern (spots) in a cubic geometry were created. The resulting $2048 \times 2048$ px images were then scaled and summed to a single 2D image, which was used to evaluate the effects of various parameters in IsoDAT2D. In the following, the settings and quality of identified data from varying mask geometry, number of datasets, and data point resolution.

### S1.2 Mask Geometry

Initial experimentation suggests that mask shape and pixel width size of mask slices are important in identifying the signal of interest without anisotropic features (Fig. S*4). The datasets for each mask were used in identical IsoDAT2D workflows, consisting of 10 iterations using the same NMF beta-divergence minimum and number of clusters (10). For data using each mask, components selected from clustering were averaged and smoothed uniformly. The simulated scattering signal of only the film, free from single crystal contributions, was used to qualitatively compare the results obtained from the IsoDAT2D process.

All mask shapes are able to identify isotropic scattering peaks in appropriate positions in $Q$; however, the relative intensities, noise, and the degree of residual single crystal signal varied with mask geometry. However, the masks with 8 sectors on one side (Fig. S*5c) result in the best match of peak shape and intensity for the IsoDAT2D-identified signal. For

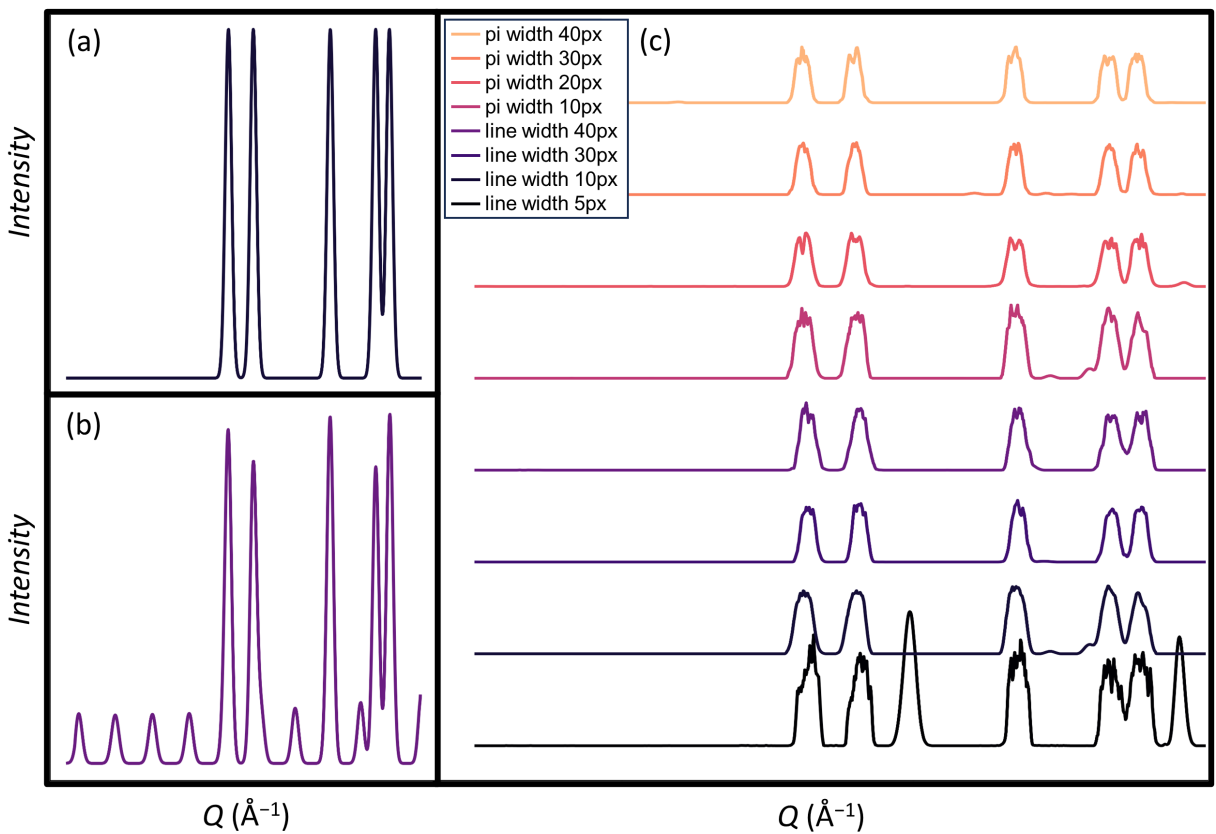

Figure S1: Analysis of mask geometry on synthetically generated data using 360 datasets. (a) The isotropic signal of interest without anisotropic effects. (b) The isotropic and the anisotropic scattering signals. (c) Outputted results after inputting (b) into the IsoDAT2D program using two mask shapes, (Fig. S*5 a,b) with four different widths for integration.

the line mask (Fig. S*5a), both widths worked similarly. However, there are unphysical deviations near the peaks' bases. Similar noise results are also apparent for the thicker, single-sector mask (Fig. S*5b). Howevr, the mirrored multi-sector mask (Fig. S*5d), and thinner single-sector mask do especially poorly with residual noise effects.

3

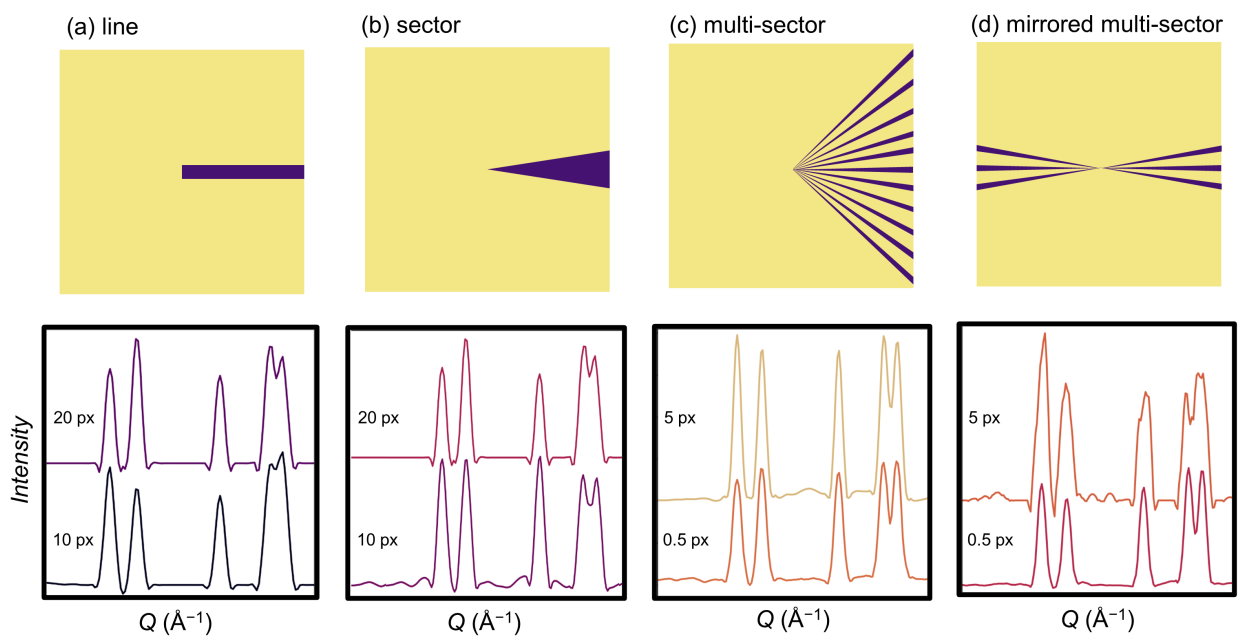

Figure S2: Experimental results of varying integration masking shapes on SimDAT2D generated thin film data. (a) Rectangular-shaped *line* mask with IsoDAT2D-identified components using 20 (top) and 10 (bottom) pixels width. (b) Sector-shaped mask with IsoDAT2D-identified components using 20 (top) and 10 (bottom) pixel widths. (c) Multi-sector mask with IsoDAT2D-identified components using sector widths of 5 (top) and 0.5 (bottom) pixels, and (d) mirrored sector slices with IsoDAT2D-identified components using sector widths of 5 (top) and 0.5 (bottom).

## S1.3 Number of datasets

The 'multi-mask' approach increases the number of datasets from 360 1D datasets by using 2 or more masks that are combined, such that the total number of datasets is the product of 360 and the number of masks, ($360 \times N_m = N_d$, where $N_m$ = number of masks and $N_d$ = number of datasets). We use five masks, specifically with 1, 2, 3, 4, and 5 sectors, each with a width of 0.5 px and an offset between sectors of 5 px. Compared with initial experiments of using 360 datasets from one masked 2D image (Fig. S*1), using 1800 datasets from one 2D image with five unique masks results in clusters that contain the isotropic scattering signal without any noise (Fig. S*2).

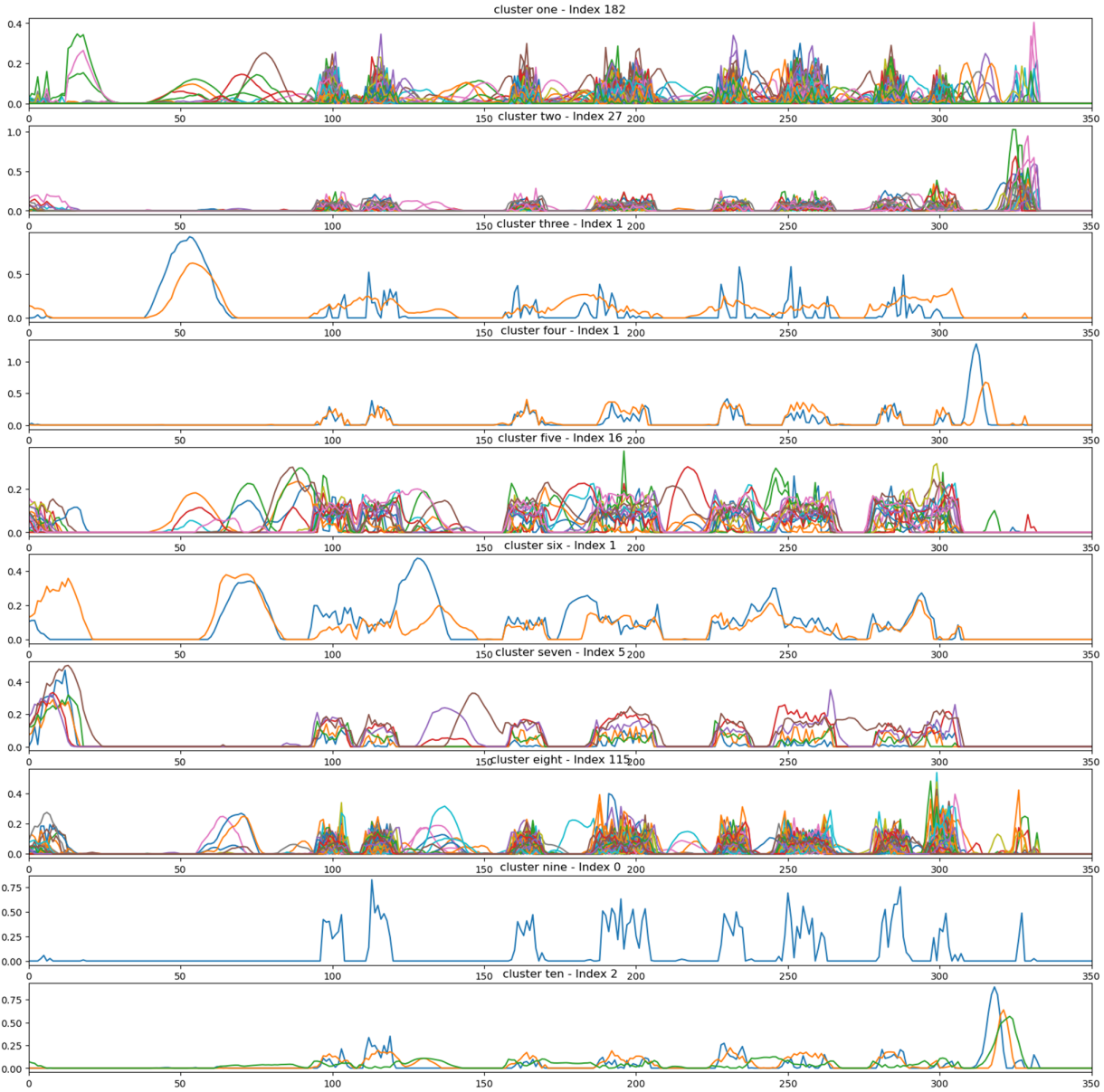

Figure S3: Direct output of the agglomerative clustering from the IsoDAT2D program using a single mask (360 datasets) made by using 5 slices at 0.5 pxwidth offset at 5 rotational degrees apart from one another. 10 clusters were used where index defines how many NMF components are in an individual cluster. Many clusters have identified a synthetically generated isotropic signal. However, there are high processing noise effects that require post-processing methods, such as smoothing.

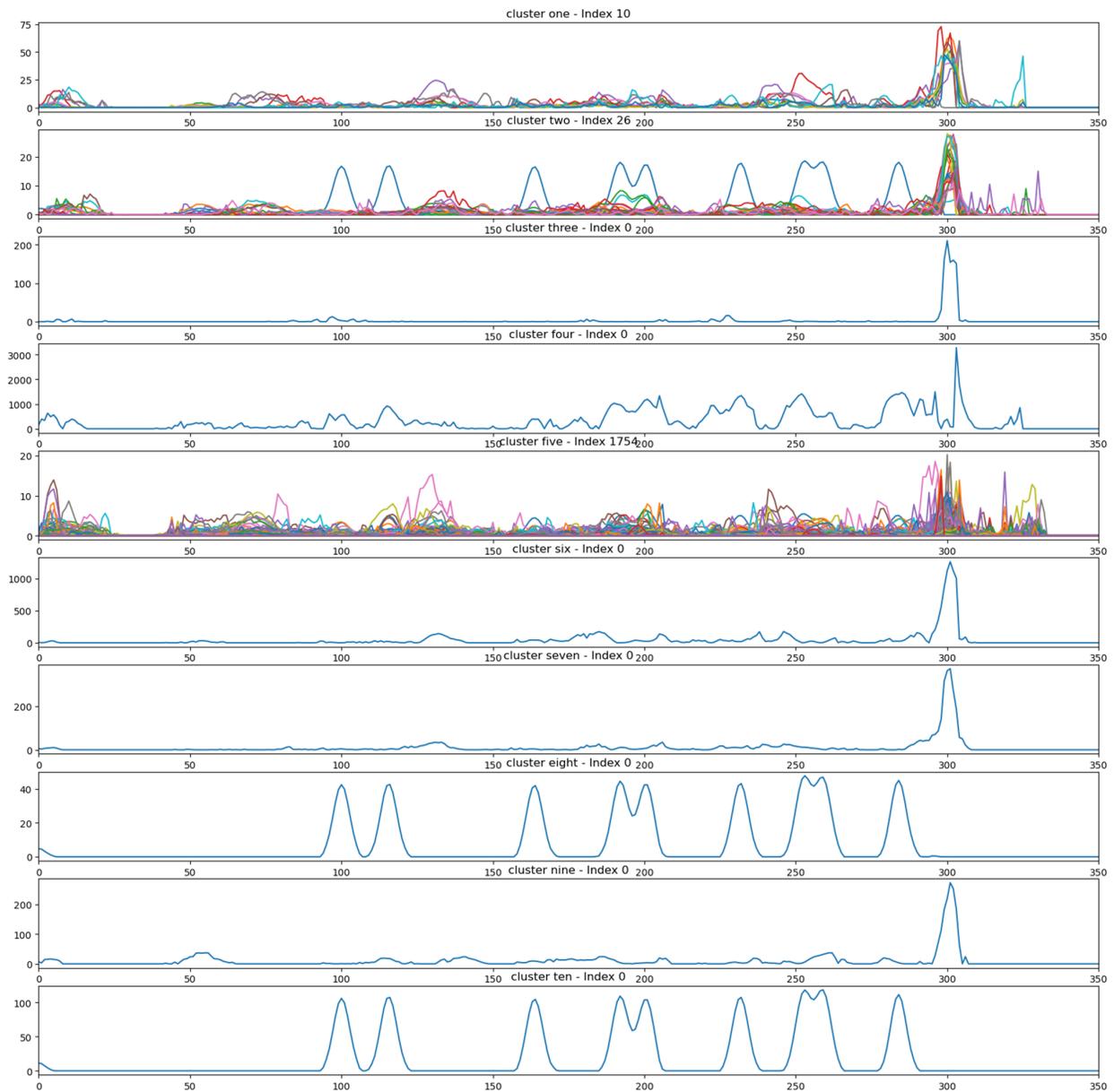

Figure S4: Direct output of the agglomerative clustering step from the IsoDAT2D Python module using multiple masks (5) having up to 5 sector slices (1800 datasets) with 0.5 px width and an offset of 5 degrees of rotation from one another. 10 clusters were created, where the isotropic signal of interest can be identified prominently without need for post-processing smoothing.

Using this larger number of 1D integrations, on the order of thousands rather than hundreds, increases the redundancy of features inputted into the NMF algorithm, which improves the performance in decomposing the data into unique types of features (Fig. S * 3). Compared to the performance of IsoDAT2D using any single mask, the quality of the signal identified with 5 uniquely generated masks was greatly improved. Specifically, the isolated signal from the multi-mask approach has features only from the isotropic signal, and with lowered processing noise, such that filtering and smoothing are not needed. We speculate that the observed noise reduction is analogous to the statistical benefits of azimuthal integration to data quality in typical total scattering data processing.

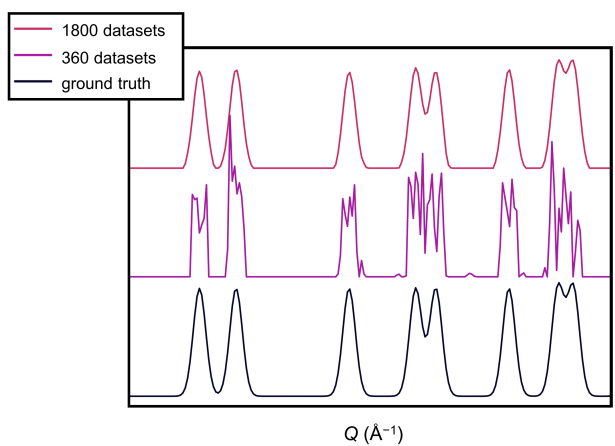

Figure S5: A SimDAT2D synthetic 1D integration from a $CeO_2$ thin film of only isotropic scattering (black) is used as a ground truth to compare IsoDAT2D-identified signals from isotropic and anisotropic scattering. The addition of 1440 more datasets (red) inputted to the program compared to 360 datasets (magenta) improved the identification of the isotropic signal of interest and reduced the amount of processing noise in the identified signal.

## S1.4 Noise Mitigation

To understand the effects of noise on IsoDAT2D behavior, we incorporate different noise intensities into synthetic 2D data. As more noise is added to the 2D data, the signal-to-noise ratios in the corresponding 1D linescans decrease (Fig. S*6). These integrated data highlight the difficulty of distinguishing the background, *i.e.*, substrate and noise, from the signal of interest. This challenge to differentiate scattering signals manifests as a lack of redundancy of features in NMF and, accordingly, hinders the identification of data features from the film and the substrate.

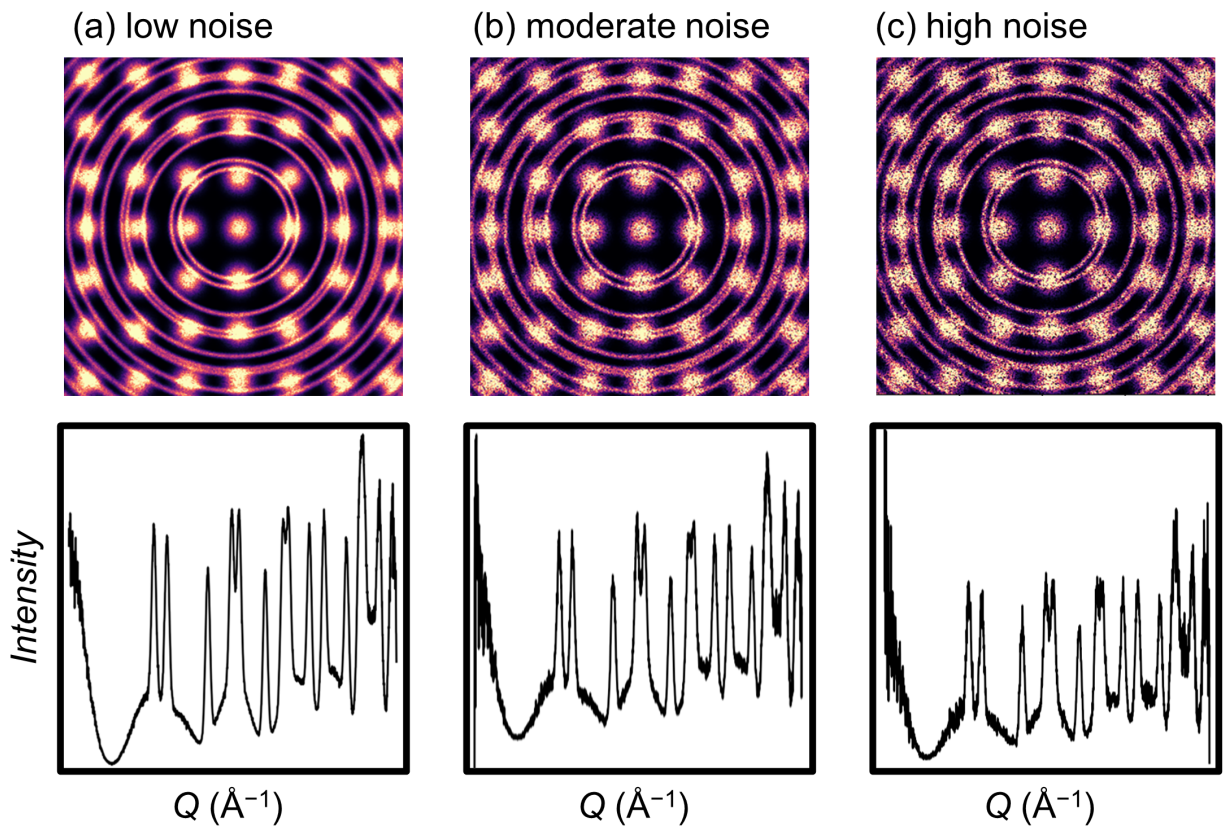

Figure S6: SimDAT2D synthetic 2D images of a thin film deposited on a single crystal substrate with varying levels of applied noise: (a) low, (b) moderate, and (c) high, along with their corresponding 1D integrations. Where low is 2, moderate is 5, and high is 7.5 standard deviations within the original data value.

In initial experiments to mitigate noise effects, we found a correlation between integration data point resolution and the degree of noise using 360 1D linescans (Fig. S*7).

However, decreasing the integration resolution is associated with a loss of information. To balance low noise without compromising information, we applied a moderate noise filter, using a random normalized distribution of 5 standard deviations, to identical integration and IsoDAT2D input parameters for 1D linescans generated using a variety of resolutions (*i.e.*, 200, 400, 800) to bin the data for less noisy results. For the 1000 data point resolution dataset, the default parameter for integration, a pre-processing Savitzky-Golay filter was applied to mitigate noise effects (Fig. S*8).

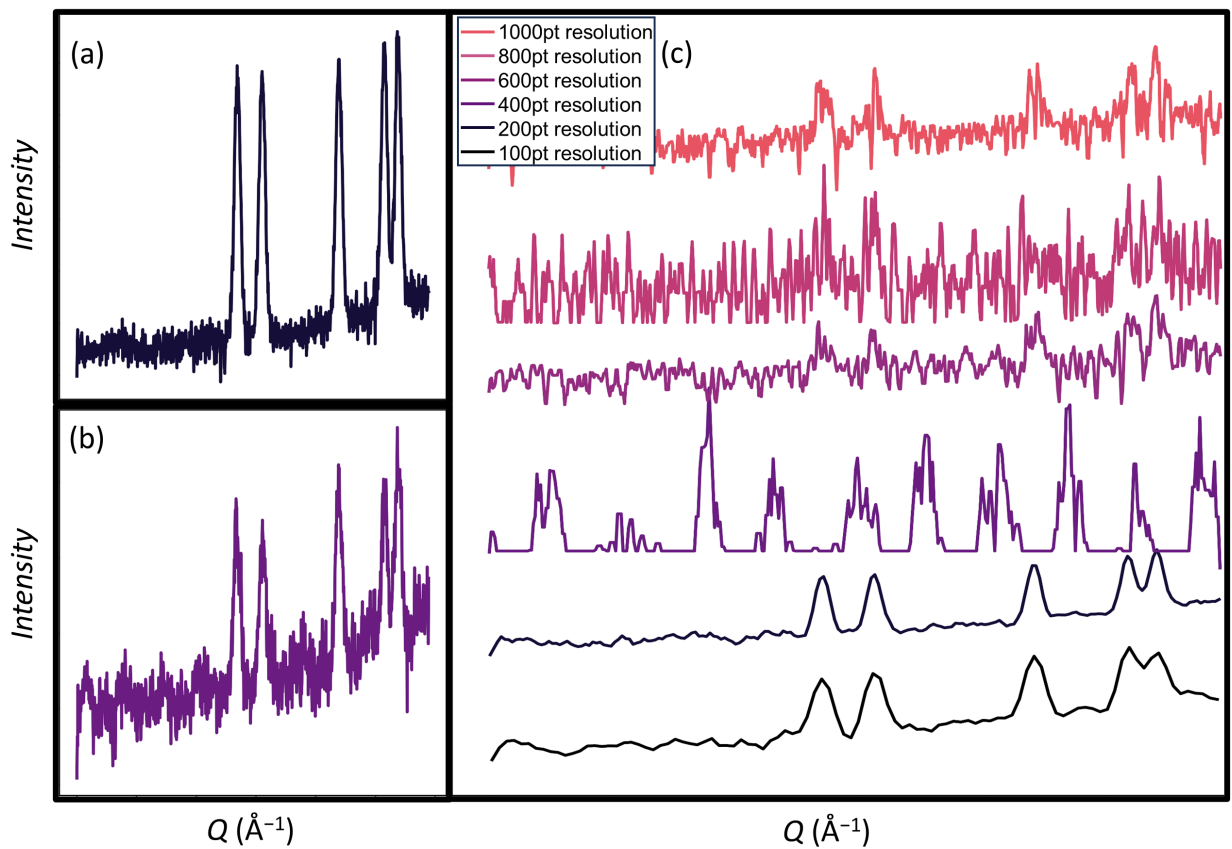

Figure S7: Analysis of non-normalized noise on synthetically generated data using 360 datasets. (a) The noisy isotropic signal of interest without anisotropic effects. (b) The noisy isotropic and the anisotropic scattering signals. (c) Outputted results after inputting (b) into the IsoDAT2D program using various data point resolutions for integration.

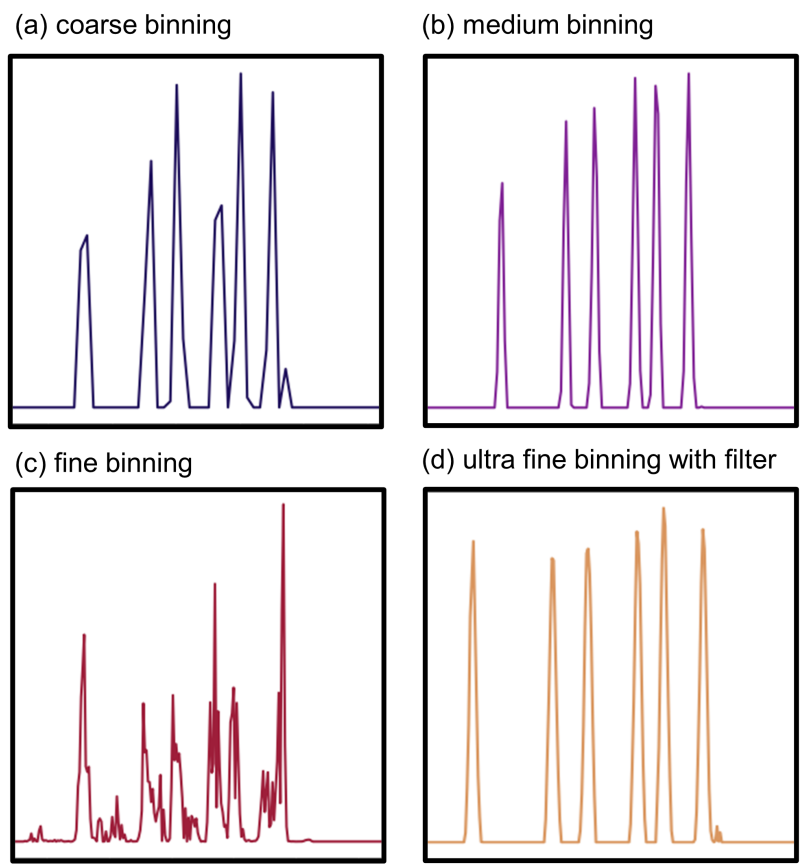

Figure S8: A moderately noisy SimDAT2D synthetic 2D scattering signal (Fig.,S*7) was used to identify a preprocessing technique that mitigates measurement noise while maintaining isotropic scattering features. The IsoDAT2D-identified $I(Q)$ isotropic scattering signals were preprocessed with different binning resolutions: (a) 200, (b) 400, and (c) 800 data points, as well as (d) filtered using a Savitzky-Golay filter with 1000 data points.

## S2  Experimental Thin Film Data

### S2.1  Methods

Two 300 nm thick GST films were deposited onto 500 $\mu$m thick substrates, one on single crystal Si (100) substrate and one on amorphous silica. Films were deposited using direct current magnetron sputtering. X-ray total scattering data were collected at the National Synchrotron Light Source II (NSLS-II) at Brookhaven National Laboratory. Data were collected using a PerkinElmer XRD 1621 detector with a sample-to-detector distance of 250 mm and X-rays with wavelength 0.1667 Å (74 keV). A Ni standard was used to calibrate the measurement geometries for data reduction.

### S2.2  Substrate Subtraction

The two GST films were processed using the same substrate subtraction technique (Fig. S*9), where measurements are taken for both the substrate alone and the film on the substrate. Subtracting the amorphous substrate from the GST film data resulted in an $I(Q)$ signal that resembles isotropic scattering. In contrast, subtracting the single crystal substrate produced positive and negative peaks due to the significant Si scattering signal remaining in the subtracted data.

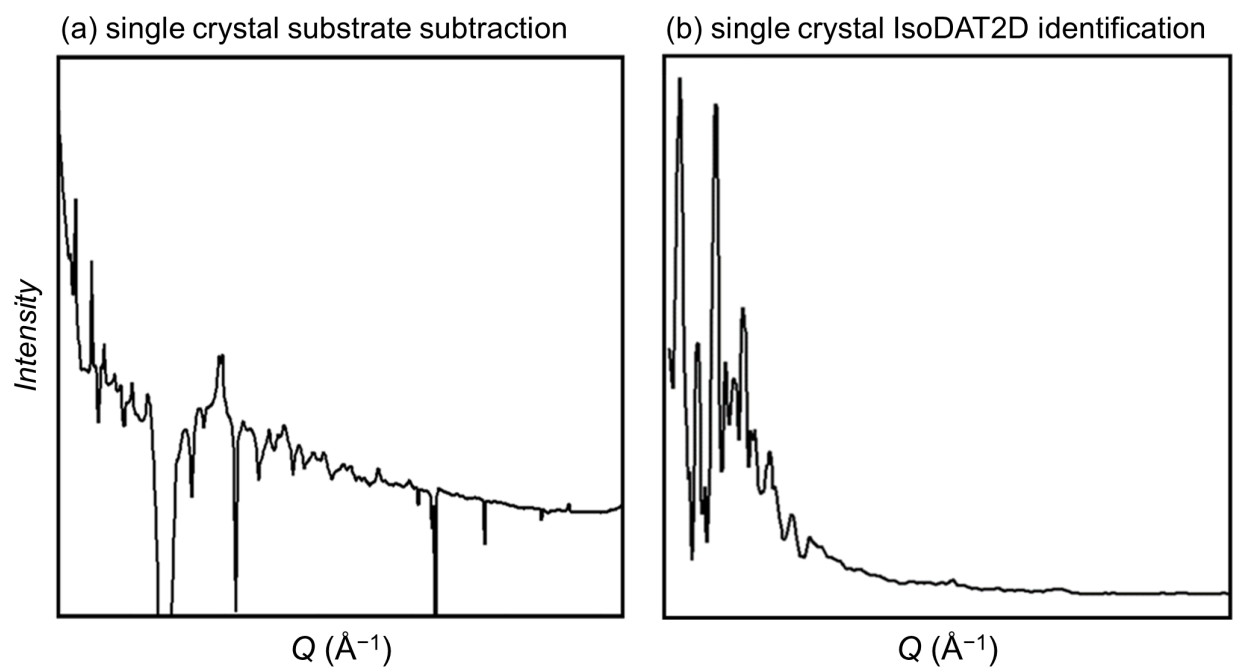

Figure S9: Comparison of the substrate subtraction and IsoDAT2D identification methods for the same thin film deposited on a single crystal total scattering measurement. (a) Application of the substrate subtraction method failed to capture the thin film scattering signal, leaving unphysical artifacts caused by high-intensity bright spots. (b) In contrast, the IsoDAT2D identification method successfully captured the thin film scattering signal, providing data suitable for PDF analysis.

13



\end{document}